\documentclass{pasj00}
\usepackage{natbib}
\usepackage{times}
\usepackage{url}

\def\th{{\rm th}}
\def\pl{{\rm pl}}
\def\bh{{\rm BH}}
\def\ff{{\rm ff}}

\begin{document}
\title{Unveiling the Nature of Coronae in Active Galactic Nuclei through Sub-mm Observations}

\author{Yoshiyuki Inoue\altaffilmark{1} and Akihiro Doi\altaffilmark{1}} 
\affil{$^1$Institute of Space and Astronautical Science JAXA, 3-1-1 Yoshinodai, Chuo-ku, Sagamihara, Kanagawa 252-5210, Japan}
\email{E-mail: yinoue@astro.isas.jaxa.jp}

\maketitle

\begin{abstract}
The heating mechanism of a corona above an accretion disk in active galactic nuclei (AGNs) is still unknown. One possible mechanism is magnetic reconnection heating requiring energy equipartition between magnetic energy and gas energy in the disk. Here, we investigate the expected observed properties in radio band from such a magnetized corona. A magnetized corona can generate synchrotron radiation since a huge amount of electrons exists. Although most of radiation would be absorbed by synchrotron self-absorption, high-frequency end of synchrotron emission can escape from a corona and  appears at the sub-mm range. If only thermal electrons exist in a corona, the expected flux from nearby Seyferts is below the Atacama Large Millimeter/ submillimeter Array (ALMA) sensitivity. However, if non-thermal electrons coexist in a corona, ALMA can measure the non-thermal tail of the synchrotron radiation from a corona. Such non-thermal population is naturally expected to exist if the corona is heated by magnetic reconnections. Future ALMA observations will directly probe the coronal magnetic field strength and the existence of non-thermal electrons in coronae of AGNs.
\end{abstract}

\section{Introduction}
\label{intro}
It is believed that X-ray radiation of radio-quiet active galactic nuclei (AGNs)\footnote{Although the term of AGNs include radio-loud population, we call radio-quiet AGNs as AGNs for simplicity in this letter.} arises from inverse Compton scattering of disk photons in an optically thin or at most moderately thick hot corona above an accretion disk \citep[see, e.g.,][]{kat76,poz77,sun80}. X-ray observations have indicated that the coronal temperature is $\sim10^9$~K \citep[e.g.][]{zdz94}. However, the heating mechanism of a corona has been a long mystery.

The accretion disk is expected to be strongly magnetized. The magneto-rotational instability would be the main mechanism of angular momentum transportation in accretion disk systems \citep{vel59,cha60,bal91}, although the detailed properties of this instability are not completely understood. Magnetic reconnection is suggested to be a plausible corona-heating mechanism on the analogy of the solar corona \citep[e.g.][]{gal79,dim98}. \cite{liu02,liu03} constructed a model for a magnetic reconnection-heated corona above an accretion disk in AGNs and Galactic black hole X-ray binaries (XRBs). The magnetic reconnection-heated corona model well reproduced the observed X-ray properties of AGNs and XRBs under the assumption of the equipartition of magnetic energy with gas energy in the disk. 

Since non-thermal electrons are observed in solar flares \citep{shi95,lin05} and the Earth's magnetotail \citep{oie02}, where magnetic reconnection accelerates particles, non-thermal electrons would exist in a reconnection-heated corona together with thermal electrons. Interestingly, such non-thermal electrons produce a power-law tail in the MeV band of AGNs' spectra, although MeV power-law tails have never been observationally confirmed. If the spectral index of non-thermal electron distribution is $\sim4$, which is similar to the measured index in solar flares \citep{lin05} and the magnetotail \citep{oie02}, the cosmic MeV gamma-ray background radiation can be naturally explained by AGNs of which the cosmic X-ray background radiation is composed \citep{ino08}. XRBs are also accretion disk systems, although the central black hole mass is of solar mass size. In fact, X-ray spectra of Seyferts resemble those of XRBs \citep{zdz99}. Observationally, it is known that XRBs extend their X-ray emission to the MeV region with a power-law \citep{mcc94,gie99}.

Here, if a corona is strongly magnetized and has a huge amount of electrons, synchrotron radiation is generated \citep[e.g.][]{ryb79,jon79,pet81,tak82,pet83,rob84,war00,oze00,war01}. In the hot accretion flow, so-called advection-dominated accretion flow (ADAF), synchrotron emission is a key ingredient. For example, the radio--infrared spectrum of Sgr A$^*$ is well reproduced by synchrotron emission based on ADAF models \citep[e.g.][]{nar95,yua03,yua04}. AGNs are believed to possess a standard accretion disk \citep[see e.g.][]{liu08}. A standard accretion disk is a radiatively efficient, geometrically thin, and optically thick disk \citep{sha73}. In the standard accretion disk, a corona is expected to have higher electron density and lower electron temperature. Most of synchrotron emission would be absorbed by the emitting electrons, so-called synchrotron self-absorption, because of high electron density. However, if the unabsorbed synchrotron component is observed from AGNs, we can investigate the magnetic field and the fraction of non-thermal electrons in a corona which have never been investigated with current telescopes. And, it can test the magnetic reconnection-heated corona model. 

In this letter, we investigate the expected synchrotron radiation from a AGN corona taking a nearby bright Seyfert IC 4329A as an example with the latest observational information \citep{bre14}. We also study parameter ranges of magnetic field strength and non-thermal electron fraction which will be constrained by future sub-mm observations. Throughout this letter, we adopt the standard cosmological parameters of $(h, \Omega_M , \Omega_\Lambda) = (0.7, 0.3, 0.7)$.

\section{Synchrotron Emission from a Corona}
\label{sec:for}

We basically follow the formalism by \citet{oze00} which formulated the synchrotron emission from thermal and non-thermal electrons. Although they focused on ADAFs, their formalism can be adopted to systems with isotropically distributed electrons with temperature $T_e\ge5\times10^8$~K \citep{mah96,oze00}. We briefly review their formalism in this section.

\subsection{Thermal and Non-thermal Electrons}
Thermal electrons are distributed following the relativistic Maxwell-Boltzman distribution given by
\begin{equation}
n_\th(\gamma_e)=n_{\th,0}\gamma_e^2\beta_e\exp(-\gamma_e/\theta_e)/[\theta_eK_2(1/\theta_e)],
\end{equation}
where $\gamma_e$ is the electron Lorentz factor, $n_{\th,0}$ is the number density of thermal electrons, $\beta_e$ is the relativistic electron velocity, $\theta_e\equiv k_{\rm B}T_e/m_ec^2$. $k_{\rm B}$ is the Boltzmann constant, $T_e$ is the electron temperature in a corona, $m_e$ is the electron rest mass, and $c$ is the speed of light. $K_2(1/\theta_e)$ is a modified Bessel function of second order. For the non-thermal electrons, we use a power-law distribution from $\gamma_e=1$ to infinity. The distribution of non-thermal electrons is given by
\begin{equation}
n_\pl(\gamma_e)=n_{\pl,0}(p-1) \gamma_e^{-p},
\end{equation}
where $n_{\pl,0}$ is the number density of non-thermal electrons and $p$ is the spectral index of the distribution. In this letter, we set $p=3.8$ which is required to explain the cosmic MeV gamma-ray background \citep{ino08} and also similar to that observed in solar flares \citep{lin05} and the Earth's magnetotail \citep{oie02}. The number density of non-thermal electrons is given by
\begin{equation}
n_{\pl,0}=\eta a(\theta_e)\theta_e(p-2)n_{\th,0}/(p-1),
\end{equation}
where $a(\theta_e)\approx(6+15\theta_e)/(4+5\theta_e)$ \citep{oze00}. Here $\eta$ is the energy fraction of non-thermal electrons in a corona. A corona is expected to have $\sim4$\% of the total electron energy in non-thermal electrons to explain the comic MeV gamma-ray background radiation \citep{ino08}. In this letter, we assume $\eta=4$\% unless otherwise noted. Observationally, the non-thermal fraction is constrained to be $\lesssim$15~\% from the spectrum of the brightest Seyfert~1 NGC~4151 \citep{joh97}.

\subsection{Synchrotron Emissivity}
Under the assumption of an isotropic distribution of electron velocities, the synchrotron emissivity of thermal electrons with temperatures in the range of $5\times10^8\  {\rm K} \le T_e \le 3.2 \times 10^{10} \ {\rm K}$  is approximated \citep{mah96} as
\begin{equation}
j_{\nu,\th}(\nu)=\frac{n_{\th,0}e^2}{\sqrt{3}cK_2(1/\theta_e)}\nu M(x_M),
\end{equation}
where $x_M \equiv 2\nu/3\nu_b\theta_e^2$, the cyclotron frequency $\nu_b=eB/2\pi m_e c$ where $e$ is the elementary electric charge and $B$ is the magnetic field strength and 
\begin{eqnarray}
\nonumber
M(x_M)&=&\frac{4.0505a}{x_M^{1/6}}\left(1+\frac{0.4b}{x_M^{1/4}}+\frac{0.5316c}{x_M^{1/2}}\right)\\
&\times&\exp(-1.8896x_M^{1/3}).
\end{eqnarray}
Here we set $a=0.0431$, $b=10.44$, $c=16.61$ \citep[Table. 1 in][]{mah96}, since we consider a corona with temperature of $\sim50$~keV ($\sim6\times10^8$~K) as discussed later.  In the case of ADAFs whose temperature is $\sim10^{10}$~K, these three parameters can be approximated to unity. The synchrotron emission is absorbed by the emitting electrons themselves. The absorption coefficient is given by $\alpha_{\nu,\th}(\nu)=j_{\nu,\th}(\nu)/B_\nu(\nu,T_e)$, where $B_\nu(\nu,T_e)$ is the blackbody source function.

For power-law distributed non-thermal electrons, the synchrotron emissivity is given by \citep{ryb79}
\begin{equation}
j_{\nu,\pl}(\nu)=C_\pl^j \eta \frac{e^2n_{\th,0}}{c}a(\theta_e)\theta_e\nu_b\left(\frac{\nu}{\nu_b}\right)^{(1-p)/2},
\end{equation}
where 
\begin{eqnarray}
\nonumber
C_\pl^j&=&\frac{\sqrt{\pi}3^{p/2}}{4}\frac{(p-1)(p-2)}{(p+1)}\\
&\times& \frac{\Gamma(p/4+19/12)\Gamma(p/4-1/12)\Gamma(p/4+5/4)}{\Gamma(p/4+7/4)}.
\end{eqnarray}
$\Gamma$ represents the Gamma function. The absorption coefficient is 
\begin{equation}
\label{eq:alpha_pl}
\alpha_{\nu,\pl}(\nu)=C_{\pl}^{\alpha}\eta\frac{e^2n_{\th,0}}{c}a(\theta_e)\theta_e\left(\frac{\nu_b}{\nu}\right)^{(p+3)/2}\nu^{-1},
\end{equation}
where
\begin{eqnarray}
\nonumber
C_\pl^\alpha&=&\frac{\sqrt{3\pi}3^{p/2}}{8}\frac{(p-1)(p-2)}{m_e}\\
&\times& \frac{\Gamma(p/4+1/6)\Gamma(p/4+11/6)\Gamma(p/4+3/2)}{\Gamma(p/4+2)}.
\end{eqnarray}

For simplicity, we assume a homogeneous spherical corona around the black hole. In this case, the observed luminosity is estimated as \citep[see sec 7.8 in][]{der09}
\begin{equation}
L_{\nu}(\nu)= j_\nu(\nu)V_c\frac{3u(\tau_{\rm syn})}{\tau_{\rm syn}},
\label{eq:lum}
\end{equation}
where $j_\nu(\nu)=j_{\nu,\th}(\nu)+j_{\nu,\pl}(\nu)$, $V_c=4\pi r_c^3/3$ is the volume of the corona, $\tau_{\rm syn}$ is the synchrotron self-absorption optical depth given as $\tau_{\rm syn}=2(\alpha_{\nu,\th}+\alpha_{\nu,\pl})r_c$, and 
\begin{equation}
u(\tau_{\rm syn})=\frac{1}{2}+\frac{\exp(-\tau_{\rm syn})}{\tau_{\rm syn}}-\frac{1-\exp(-\tau_{\rm syn})}{\tau_{\rm syn}^2}.
\end{equation}
We set the size of the corona is $r_c=10R_s$ where $R_s$ is the Schwarzschild radius \citep[e.g.][]{liu02}.

When a Compton {\it y}-parameter $y\equiv(4k_{\rm B}T_e/m_ec^2)/\tau_{T}$ is an order of unity where $\tau_T$ is the Thomson scattering optical depth, the gas density of a hot corona is $n_{\th,0}=\tau_T/(\sigma_Tl)\simeq2\times10^9\ {\rm cm}^{-3}$ assuming the case with a black hole mass of $10^8M_\odot$ and $T_e=5\times10^8$~K. Here $\sigma_T$ is the Thomson scattering cross section and we assume $n_{\th,0}\gg n_{\pl,0}$.
Under the assumption of energy equipartition between magnetic field and gas in the disk, the magnetic field strength becomes $10^3$~G with the typical radiation efficiency of 10~\%.

\section{Case for IC~4329A}
As an example, we consider IC~4329A which is a nearby bright Seyfert 1.2 galaxy in the southern hemisphere. IC~4329A locates at the distance of redshift $z=0.0161$ \citep{wil91} and its central supermassive black hole mass $M_{\bh}\simeq1.20\times10^8M_\odot$ \citep{del10}. Recent joint {\it NuSTAR}/{\it Suzaku} X-ray observation determined the properties of the corona as $k_{\rm B}T_e=50_{-3}^{+6}$~keV with $\tau_{T}=2.34_{-0.11}^{+0.16}$ in the case of spherical geometry \citep{bre14}. Following section \ref{sec:for}, we can evaluate the expected synchrotron emission from IC 4329A with these latest observational information.

\begin{figure}[t]
\centering
\includegraphics[width=8.0cm]{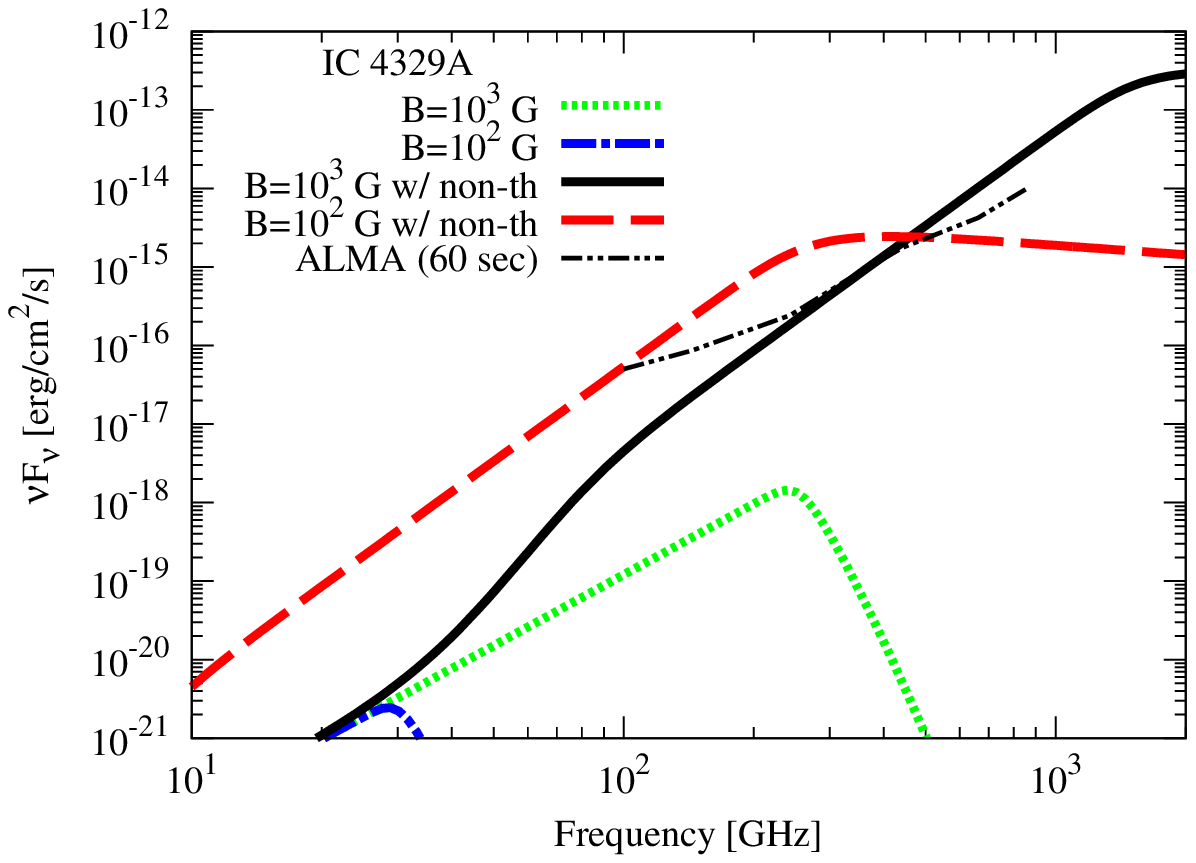} 
\caption{Predicted synchrotron spectra from the corona of IC 4329A at $z=0.0161$ \citep{wil91} assuming spherical geometry. We set $M_{\bh}=1.2\times10^8M_\odot$ \citep{del10}, $k_{\rm B}T_e=50$~keV, and $\tau_{T}=2.34$ \citep{bre14}. The fraction of non-thermal electron energy is set to be $\eta=4$\% with the spectral index of 3.8 \citep{ino08}. Dotted and dot-dashed curve represents the case for thermal electrons only with $B=10^3$~G and 10$^2$~G, respectively. Solid and dashed curve represents the case for thermal+non-thermal electrons with $B=10^3$~G and 10$^2$~G, respectively. Double dot-dashed curve represents the cycle-2 ALMA sensitivity with 60 seconds integration.\label{fig:all} }
\end{figure}

Seyfert galaxies are known to have weak jet emission in the radio band. IC 4329A also has radio emission \citep[see e.g.][]{nag99}. Very Large Array (VLA) observations indicated the (simultaneous) two-point spectral index between 1.4~GHz and 8.4~GHz is $\sim-0.96$ using the AB-array configuration. The observed flux at 1.4~GHz and 8.4~GHz is 60~mJy and 10.7~mJy, respectively \citep{nag99}. Then, the extrapolated $\nu F_\nu$ flux is $\simeq1.0\times10^{-15}\ {\rm erg \ cm^{-2} \ s^{-1}}$ in the sub-mm band. On the basis of our analyses of the NRAO VLA Sky Survey (NVSS by the D-array) and VLA archival data (observation code: AK0394 by the A-array), we obtained results of  66.8~mJy at 1.4~GHz and 11.4~mJy at 8.4~GHz, which are consistent with \citet{nag99}, indicating  a spectral index of $-0.96$ is not affected by resolution effect or flux variation significantly. NASA/IPAC Extragalactic Database (NED)\footnote{NED is operated by the Jet Propulsion Laboratory, California Institute of Technology, under contract with the National Aeronautics and Space Administration.} provides flux data at 843~MHz as 79.7~mJy, indicating a flatter radio spectrum ($\sim -0.35$) at lower frequencies. These observations suggest that synchrotron self-absorption occurring around $\sim$1~GHz corresponding to the parsec-scale radio source, i.e. the jet component. 

Dust emission from the host galaxy would also contaminate to the sub-mm emission. Thanks to high angular resolutions of the full Atacama Large Millimeter/ submillimeter Array (ALMA; e.g., $\sim0.015$~arcsec, corresponding to $\sim5$~pc at the distance of IC~4329A,  at 300~GHz (band-7)), a possible dust contribution is expected to be only $\sim 1.4\times 10^{-16}\ {\rm erg \ cm^{-2} \ s^{-1}}$, assuming a dust temperature of 30~K and an emissivity index of 1.5 \citep[c.f.][]{doi11}, which is fainter than the jet contamination. We also check emission from the dust torus which has temperature of $\sim200$~K and the size of $\sim2.8$~pc \citep{tri11}. Using an AGN infrared spectrum model \citep{mul11}, the expected contribution is an order of magnitude lower than ALMA's sensitivity at 900~GHz (band-10). Thus, contaminations from the galactic dust and the dust torus are negligible in future ALMA observations. Furthermore, the coronal emission can be discriminated definitely from dust and jet contaminations based on spectral indices.

Figure. \ref{fig:all} shows the expected radio emission from the corona of IC~4329A for different magnetic field strength ($B=10^2$~G and $10^3$~G) and for different electron populations (thermal electrons only and thermal + non-thermal electrons). We also show the sensitivity limit of ALMA cycle-2 with an integration of 60 seconds\footnote{http://alma.mtk.nao.ac.jp/e/index.html} for comparison. 

The electron density in the corona is so high that most of radio emission is absorbed by ambient electrons. However, emission at higher frequency can survive from synchrotron self-absorption. If the corona contains thermal electrons only, the expected flux is far below the ALMA sensitivity limit. If non-thermal electrons exist together with thermal electrons in the corona, the predicted flux is higher than the ALMA sensitivity limit and the jet component. This is because the observable emission are emitted by high energy non-thermal electrons. Therefore, detection of synchrotron emission by ALMA would allow us to investigate not only the properties of magnetic fields in the corona but also its non-thermal content.  Importantly, future ALMA observations will be able to see both of the corona and the jet. Hereinafter, we focus on models with thermal + non-thermal electrons.

\begin{figure}[t]
\centering
\includegraphics[width=8.0cm]{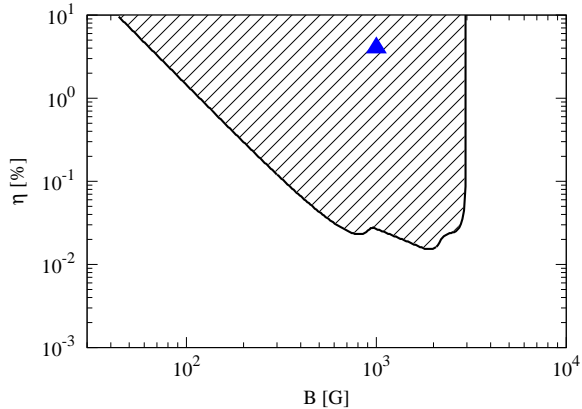} 
\caption{The testable region of the magnetic field strength $B$ and the energy fraction of non-thermal electrons $\eta$ in the corona for IC~4329A with ALMA. The shaded region represents the ALMA detectable range with the cycle-2 sensitivity.The triangle point show the value of $\eta=4$\% explaining the cosmic MeV gamma-ray background and $B=10^3$~G corresponding to the equipartition value. The other parameters are the same as in Figure. \ref{fig:all}. We take into account the galactic dust, dust torus, and jet contamination (see the text for details).  \label{fig:range} }
\end{figure}

Figure. \ref{fig:range} shows the detectable parameter ranges of magnetic field $B$ and the fraction of non-thermal component $\eta$ with ALMA. Here we take into account the galactic dust, dust torus, and jet contamination. The triangle point represents the value of $\eta=4$\% with which AGNs can explain the cosmic MeV gamma-ray background radiation and $B=10^3$~G at which magnetic energy is in energy equipartition with gas energy in the disk. The ALMA sensitivity allows us to investigate $\eta$ down to 0.02\% for $B=10^3$~G which is limited by the contamination of the jet emission. And we can investigate the coronal magnetic field strength down to $B=50$~G, corresponding to $\sim0.3$\% of gas energy in the disk, with $\eta=4$\%. Thus, wide range of $B$ and $\eta$ can be tested with ALMA. In other words, we can strongly constrain the coronal properties even if ALMA does not see the coronal synchrotron emission.

\section{Discussion and Conclusion}
In this letter, we have investigated the method to probe the magnetic field and non-thermal contents in AGN coronae. Future ALMA sub-mm observations of nearby bright AGNs such as IC~4329A will be able to probe these new physical information of the corona and test the magnetic-reconnection heated corona scenario \citep{liu02}, although it is difficult to investigate the case with a pure thermal corona. For the case of IC~4329A, we can investigate wide range of the magnetic field strength and the energy fraction of non-thermal electrons. Coronae are heated by magnetic reconnection if $B\gtrsim10^3$~G and the MeV background flux is explained by Seyferts if $\eta\simeq4$\%. 

If magnetic reconnection can not provide enough energy to heat up the corona, another corona heating mechanism is required. However, viscous heating by the coronal accretion flow is not enough to keep itself hot  against Compton cooling \citep{liu02_acc}. Moreover, thermal energy content of coronae is too low to explain the observed X-ray luminosities \citep{mer01}. 

In case of shortage of non-thermal electrons for explaining the MeV background flux, other candidates may explain it, such as type Ia supernovae \citep[SNe Ia;][]{cla75,zdz96,wat99}, blazars \citep{aje09}, and radio galaxies \citep{str76}. However, recent studies show that those are not enough to explain the MeV background flux, see \citet{hor10} for SNe la, \citet{ino09,aje12} for blazars, and \citet{mas11,ino11} for radio galaxies. Although annihilation of dark matter particles may explain the MeV background flux \citep{oli85,ahn05_dm2}, there are less natural dark matter candidates, with a mass scale of MeV energies, rather than GeV-TeV dark matter candidates.

Even if ALMA does not see the coronal synchrotron emission, we can probe the nature of the corona combining sub-mm and X-ray observations. The fraction  of the non-thermal component can be revealed through observations of MeV tails in nearby AGNs' spectra \citep{ino08} or the MeV background anisotropy by the {\it Astro-H} X-ray satellite \citep{ino13}. It will give tighter constraints on the magnetic field (see Figure. \ref{fig:range}). 
Therefore, future sub-mm and X-ray observations will be complementary with each other to probe AGN coronae.

For simplicity, we have considered a spherical AGN corona. Our method can be applied to other AGN objects, Galactic black hole systems, and different coronal geometries. Recent X-ray observational studies suggest that coronae have extended geometry like a slab rather than spherical geometry \citep[see e.g.][]{fab09,zog10,wil12,fab12,kar13}. In the slab corona case, $\tau_{T}$ becomes $0.068\pm0.02$ \citep{bre14}. Following an extended slab geometry \citep{wil12}, the resulting synchrotron self-absorption is expected to be less effective than the case in the spherical geometry as in $\tau_{T}$. Therefore, we can expect more synchrotron photons.

IC~4329A has relatively lower coronal temperature ($T_e\simeq50$~keV) than typical AGNs do ($T_e\simeq 100$~keV). The cut-off energy $E_{\rm cut}$ in AGNs' spectra is expected to be above 200~keV \citep{vas13}, which implies $T_e\gtrsim70$~keV. Fixing the Compton {\it y}-parameter, $n_e$ is proportional to $kT_e$. The expected flux becomes higher by a factor of temperature differences. We note that synchrotron self-absorption turnover frequency is less sensitive to the electron density (see Equation. \ref{eq:alpha_pl}).

How distant object can ALMA observe the coronal synchrotron emission? If there is an object with the same physical parameters as IC~4329A with $B=10^3$~G and $\eta=4$\%. We can probe coronal synchrotron emission from such AGNs up to $\sim200$~Mpc with ALMA. If far as the size of the host galaxy, the dust torus, or the jet is larger than ALMA's resolution, we would expect the same flux level from each component. ALMA can resolve galactic dust even at the distance of $\sim1$~Gpc assuming galactic dust distributing over 10~kpc. Dust torus and jet components can not be resolved at the distance above a few hundred kpc because they are pc-size. This, in turn, means that the ratio between coronal synchrotron emission and dust torus/jet emission will not change. On the contrary, the ratio between coronal emission and galactic dust emission will change with distance because the flux from galactic dust will not change \footnote{ALMA can detect the galactic dust component by shortening the baseline length. However, we consider the longest baseline array, since we are interested in the central coronal emission.}. However, since the galactic dust flux is below the ALMA's sensitivity,  the detectability strongly depends on the sensitivity rather than the contamination.

If ALMA can detect polarization from the coronal synchrotron emission, it will enable us to investigate the magnetic field structure right above the the disk. Since the magnetic field structure is an important key to understand the relativistic jet launching mechanism, such observations will give an important clue in the field of jet formation physics. A detailed study of polarimetric properties is our future work.

\bigskip
We thank useful comments from anonymous referee. We also thank K. Ichikawa, N. Isobe, S. Mineshige, K. Murase, and K. Murata for useful comments and discussions. Y.I. acknowledges support by the JAXA international top young fellowship. This research has made use of the NASA/IPAC Extragalactic Database (NED), which is operated by the Jet Propulsion Laboratory, California Institute of Technology, under contract with the National Aeronautics and Space Administration.

\end{document}